# Study of Synchronous RF Pulsing in Dual Frequency Capacitively Coupled Plasma

Abhishek Verma,[1] Shahid Rauf[1], Kallol Bera[1], D. Sydorenko[2], A. Khrabrov,[3] and I. Kaganovich[3]

[1]Applied Materials, Inc., 3333 Scott Blvd., Santa Clara, California 95054, USA

[2]University of Alberta, Edmonton, Alberta T6G 2E9, Canada

[3]Princeton Plasma Physics Laboratory, 100 Stellarator Rd., Princeton, New Jersey 08543, USA

## Abstract

Low-pressure multi-frequency capacitively coupled plasmas are used for numerous etch and deposition applications in the semiconductor industry. Pulsing of the radio-frequency (RF) sources enables control of neutral and charged species in the plasma on a millisecond timescale. The synchronous (i.e., simultaneous, in-phase) pulsing of both power sources in a dual frequency capacitively coupled plasma is examined in this article. Due to the low gas pressure, modeling has been done using the electrostatic Particle-in-cell/Monte Carlo collision method. The objective of this work is to investigate the sensitivity of the plasma properties to small changes in timing during synchronous pulsing of the 2 RF sources. It is demonstrated that small deviations in the on and off times of the 2 RF sources can lead to major changes in the plasma characteristics. This high sensitivity is of concern for process repeatability but can be utilized to enable better control of the dynamics of plasma-surface interaction. In the simulations, the pulsing parameters (on and off times and ramp rates) are varied and the temporal evolution of plasma characteristics such as electron density ($n_e$), species current at the electrode, and electron temperature are examined. It is demonstrated that if the low-frequency (LF) source is turned off a few µs before (or after) the high-frequency source, $n_e$ during the off-state is significantly higher (or lower) due to the frequency coupling effect. Similarly, turning on the LF source with a small delay results in a sharp increase in the plasma density when the RF sources are turned on.

# 1. INTRODUCTION

Multi-frequency capacitively coupled plasmas (CCP) are widely used for materials processing in the semiconductor industry. [1] The combination of a high and low radio frequency (RF) source allows one to control the ion and neutral species concentrations as well as the ion energy and angular distribution (IEADF). The semiconductor industry is rapidly progressing towards sub-5 nm technology nodes where devices need to be fabricated with atomic precision. [2-4] 3-dimensional devices have also become prevalent in both memory and logic integrated circuits, pushing the plasma processing technologies to their limit. [5-6] The pulsing of RF power has proven to be an effective means of more precise control of the plasma characteristics, which is critical for the fabrication of 3D and small-scale devices. With several RF sources in multi-frequency CCPs, many different pulsing modes have been developed. This article focuses on synchronous pulsing in a dual-frequency CCP where both the high-frequency (HF) and low-frequency (LF) sources are turned on and off simultaneously. However, we introduce small delays between the turn on or off times of the two RF sources. It is demonstrated that the plasma properties are highly sensitive to these small delays. One can use these delays to control important plasma properties more precisely in pulsed plasmas. This sensitivity introduces challenges for chamber matching and process repeatability, which need to be considered while designing pulsed plasma processes.

The pulsing of CCPs has been the subject of multiple studies in the literature. Agarwal, Rauf, and Collins demonstrated that the relative phase between HF and LF pulsing can be used to control plasma uniformity in a dual-frequency CCP. [7] In a separate study, these authors examined synchronous pulsing in electronegative CCP discharges. [8] They found that applying a DC voltage on a counter electrode can enhance negative ion flux during the power-off phase. Song and Kushner used a 2-dimensional fluid plasma model to examine how the DC blocking capacitor in a dual-frequency CCP can be used to control the ion energy distribution during the power-off phase. [9] Either the LF or HF voltage was pulsed in their simulations, but not synchronously. Song and Kushner further investigated electron energy distribution (EED) during HF pulsing. [10] They found pulse repetition frequency and duty cycle to be effective knobs for controlling the EED. Wen et al. used a particle-in-cell/Monte Carlo collision (PIC/MCC) model to examine phase modulation in pulsed DF-CCP. [11] Pulsing was done at the frequency of the LF source, and either the LF or HF voltage was pulsed. It was found that higher harmonics can be excited during the initial stage of the pulse period, and these harmonics can be tuned using the phase shift. Sirse et al. did time-resolved measurements of electron density ($n_e$) in a DF-CCP where the HF was pulsed while the LF power was kept constant. [12] They found that when HF is turned off, $n_e$ decay rate is slow in Ar compared to Ar/CF$_4$/O$_2$ and faster with higher LF power. Jeon et al. examined the effect of pulse phase lag during synchronous pulsing in a DF-CCP on the etch characteristics. [13] The etch profile and etch rate were a sensitive function of phase lag, with the etch rate highest and etch most anisotropic at 0° phase lag. Kim et al. used embedded RF pulsing during the pulsing of DF-CCP where the HF duty cycle (DC) was fixed at 50% while the HF DC was varied between 30 – 90%. [14] The etch selectivity was found to be sensitive to the HF source DC. Zhang et al. discussed the control of ion energy distribution function (IEDF) using the phase shift between the first and second harmonic in a dual or triple-frequency CCP. [15] The phase influenced both the width and energy of the IEDF. A third frequency could be used to enhance the plasma density. Hernandez, Overzet, and Goeckner used phase-resolved optical emission spectroscopy to examine pulsing in a single-frequency CCP. [16] They specifically reported on the early stages of plasma reignition in an Ar discharge and found that the electron heating mode during this stage depends on the pulsing frequency. Cho et al. used several diagnostics to understand $SiO_2$ etching in pulse-modulated single-frequency CCP. [17] They focused on the pulse on and off times instead of pulsing frequency and DC. They attributed the reduction in micro-trenching with increasing on and off times to negative ion neutralization on the wafer surface. Ma et al. used a 1D PIC/MCC model to look in detail at the temporal evolution of the plasma in a dual-level synchronously-pulsed CCP. [18] Both sources were pulsed at 20 kHz, the LF was turned on and off with 50% DC, while the HF was pulsed between 2 levels. The plasma evolved quite differently during the initial stage depending on the second level. Wang et al. used a combination of experimental diagnostics, PIC/MCC modeling, and an analytical model to examine the effect of the pulse-off period on the ignition phase of a pulsed CCP operating at 450 mTorr. [19] Plasma and electrical parameters were observed to be sensitive to the afterglow period due to its impact on the charge density at the beginning of the pulse.

This article is organized in the following manner. The computational model is described in Sec. 2. The computational results are presented in Sec. 3 and a brief summary is included in Sec. 4.

## 2. COMPUTATIONAL MODEL

The simulations in this article have been done using EDIPIC, a 2-dimensional (2d3v) electrostatic particle-in-cell (PIC) modeling code. EDIPIC is open source and available on GitHub with the necessary documentation [20]. It uses a standard explicit leap-frog algorithm in Cartesian geometry, with the Boris scheme for particle advance [21]. The electrostatic field is obtained from the Poisson's equation solved using the PETSc library [22]. The code includes a Monte-Carlo model of elastic, inelastic, and ionization electron-neutral collisions. Cross-sections for collisions with Argon neutrals used in the simulations described below are from Ref. [23]. EDIPIC can simulate crucial atomistic and plasma-surface interaction processes needed for simulations of partially ionized plasmas, including, but not limited to, the secondary electron emission induced by electrons and ions. EDIPIC has been verified in several international benchmarks [24, 25].

EDIPIC is written in Fortran 90 and parallelized using Message Passing Interface (MPI). Good scalability of up to 400 CPU cores has been demonstrated. The code is equipped with numerous diagnostic capabilities, including but not limited to, the phase-space data and ion and electron velocity distribution functions, as well as spectral analysis procedures required to study wave propagation in plasmas.

Our simulations are applied to an argon discharge with a gap of 4 cm between the two electrodes. The HF pulsed voltage source is connected with the electrode at x = 0 cm and the LF pulse electrode source is at x = 4 cm, as shown in Figure 1. The HF electrode operating at 50 MHz has an applied RF potential of $V_{HF}$ = 200 V and the LF source is operating at 1 MHz with $V_{LF}$ = 750 V amplitude voltage. As is typical for dual frequency CCPs, $V_{LF} > V_{HF}$ as the HF source can produce the plasma more efficiently. The initial electron temperature, ion temperature, and density are 1 eV, 0.5 eV and $5\times10^{15}$ m$^{-3}$, respectively. The cell size in x-direction is 1/4 times the initial Debye length and is equal to 0.125 mm. The simulations have been done with 40 cells in the y direction with cell size of 0.125 mm and periodic boundary condition in the y-direction. The results have been averaged in the y-direction for the plots in this article. The time step satisfies the stability criterion $\omega_{pe}\Delta t<0.2$, where $\omega_{pe}$ is the plasma frequency. In all simulations, we first model the plasma with both voltage sources on at the given conditions for 200 µs. The plasma properties reach steady state during this time. The plasma is subsequently pulsed according to the pulsing scheme for each simulation.

The discharge is sustained at 20 mTorr in Argon gas. Elastic, excitation, and ionization collisions for electrons and elastic collisions for Ar ions are considered in the Monte Carlo collision model. Multistep ionization and metastable atoms are not included in our PIC/MCC model as they are not considered significant at low gas pressures. [26] The argon ions and electrons are assumed to be perfectly absorbed at the electrodes.

## 3. COMPUTATIONAL RESULTS

All the pulsing results presented here are for a pulsing frequency of 10 kHz and a duty cycle of 40% for synchronous pulsing. The different pulsing scenarios considered are summarized in Table 1 and sketched in Figure 2. The times, $\tau_{ramp}$, $\Delta\tau_{off}$, and $\Delta\tau_{on}$, are with reference to the time when the voltages start to change. When ramped, both voltages are linearly increased or decreased during $\tau_{ramp}$ time. In the staggered pulsing cases, the HF voltage pulsing times are not changed. $\Delta\tau_{off}$ is the difference between the times when the LF and HF voltages start to decrease from their peak level. $\Delta\tau_{on}$ is the difference between the times when the LF and HF voltages start to increase from 0 V.

We first present results for the baseline process (synchronous pulsing) with and without voltage ramping during the pulse off and on transitions. The spatiotemporal distribution of electron density ($n_e$), electron temperature ($T_e$), and plasma potential are plotted in Fig. 3 over a pulse period. When both voltage sources are turned off at 10 μs, $n_e$ starts to decrease quickly. Even more quickly, as there is no externally applied voltage, the sheaths contract and the electrons expand into the region previously occupied by the sheaths. When the HF and LF voltages are turned on at 50 μs, the sheaths initially expand quickly due to the applied RF voltage. The sheaths then slowly contract during the voltage-on phase due to plasma generation and an increase in the plasma density.

It is easier to visualize the rate of change of plasma properties by examining the spatially-averaged plasma properties. The temporal evolution of the spatially averaged $n_e$ for the base case (synchronous pulsing) with and without ramping is shown in Fig. 4 for one pulse period. Both RF voltages are turned off at 10 μs and turned on at 50 μs, but ramped in 2 μs in one of these cases. The spatially averaged electron density $n_e$ drops exponentially during the voltage off-phase due to the wall losses and lack of plasma generation (ionization). The electron temperature $T_e$ decreases even more quickly than $n_e$, which is consistent with the previous results for the power-off state [27] When both the HF and LF voltages are turned on at 50 µs, the spatially-averaged electron density initially drops to its lowest value shortly after the voltages are turned on. This drop is due to electrons that are ejected from the region near the electrodes where the sheaths form. From there, the density recovers and gradually reaches the pre-pulse level. The spatially-averaged $T_e$ is shown in Fig. 4, $T_e$ drops to its minimum value just before the pulses are turned on. After the voltages are turned on, $T_e$ initially spikes beyond the steady state value of 1.5 eV. This spike in $T_e$ is well documented in the literature. [27, 28] $T_e$ then settles to the steady-state. Voltage ramping during the off and on transitions has minimal effect on the plasma properties.

The electrons and $Ar^+$ ions are primarily lost at the walls. We have shown the electron and ion currents at the electrode at x = 0 in Fig. 5 for synchronous pulsing. When the voltages are turned off at 10 µs, there is a sharp decrease in charged species current at the electrode. This decrease in current is attributed to the sharp reduction in sheath electric field which was previously accelerating the ions and enhancing ion loss. When the voltages are turned on at 50 µs, the ion current spikes quickly before it settles to the steady-state value. The spike is due to ions that are removed from the sheath region.

Table 1: Voltage on, on, and ramp times used in the different pulsing cases

| Case Name | $\tau_{ramp}$ | $\Delta\tau_{off}$ | $\Delta\tau_{on}$ |
|---|---|---|---|
| Base case | 0 | 0 | 0 |
| Base case with ramp | 2µs | 0 | 0 |
| St-A-2.5 | 0 | -2.5µs | 0 |
| St-A-5.0 | 0 | -5µs | 0 |
| St-B-2.5 | 0 | +2.5µs | 0 |
| St-B-5.0 | 0 | +5µs | 0 |
| St-C-2.5 | 0 | 0 | -2.5µs |
| St-C-5.0 | 0 | 0 | -5µs |
| St-D-2.5 | 0 | 0 | +2.5µs |
| St-D-5.0 | 0 | 0 | +5µs |

The first variation of synchronous pulsing we consider is when the LF voltage is turned off earlier than the HF voltage, $\Delta\tau_{off} < 0$. Fig. 6 shows a comparison of spatially-averaged electron density as $\Delta\tau_{off}$ is varied between 0 and -5 µs. $n_e$ sharply increases in the short time when the LF voltage has been turned off while the HF is still on. Excess $n_e$ increases monotonically with $|\Delta\tau_{off}|$ for the time delay considered. This phenomenon has been described in the literature in the context of frequency coupling. [29] With the LF off, the sheath is thinner and HF voltage can couple more effectively to the electrons producing more electrons and ions through stochastic heating. For the short time the LF is off while the HF is still on, the electrons are more energetic, but $T_e$ quickly decreases when the HF voltage is turned off.

The effect of negative $\Delta\tau_{off}$ on temporally and spatially resolved $n_e$ is shown in Fig. 7. One can observe that during the short time the LF voltage has been turned off while the HF voltage is on, $n_e$ increases in the bulk plasma indicating that the density increase is happening due to enhanced production and not decreased loss. One can also observe that the sheath becomes thinner during this period.

We have shown the electron and ion current to the electrode at x=0 in Fig. 8 when $\Delta\tau_{off}$ = -5 μs. Comparing these results to the corresponding results for synchronous pulsing in Fig. 5, we can observe that the currents to the wall first decrease when the LF voltage is turned off and then increase until the HF voltage is turned off as well. The initial decrease after the LF is turned off indicates that LF in the sheath region was enhancing charged species loss to the walls prior to the LF being turned off. The subsequent increase is due to the overall plasma density increasing.

The next pulsing configuration we consider is where the LF voltage is turned off later than the HF voltage, $\Delta\tau_{off} > 0$. The spatially averaged electron density and temperature are shown in Fig. 9 as a function of $\Delta\tau_{off}$. $n_e$ decreases rapidly when the HF voltage is turned off while the LF voltage is still on. This sharp decrease in $n_e$ is attributed to the LF not being efficient in plasma production. The electrons and ions still get drained out of the plasma at a high rate as long as the LF voltage is on, which is discussed in more detail later. One can control $n_e$ during the voltage-off phase using positive $\Delta\tau_{off}$, but not as strongly as when $\Delta\tau_{off}$ is negative. The electron temperature results show that when the HF voltage is turned off while the LF voltage is still on, $T_e$ sharply decreases, which is due to the high energy electrons leaving more quickly in the absence of plasma production.

In Fig. 10, we show how $n_e$ varies with time and space when $\Delta\tau_{off} = 2.5$ and 5 µs. It can be seen that when the HF voltage is turned off while the LF voltage is still on, the sheaths get thicker and the bulk electron density decreases, indicating that the losses are more significant than plasma production during this phase. Fig. 11 shows the wall current when $\Delta\tau_{off} = 5$ µs. Comparing to the corresponding results for the base case (Fig. 5), the electron and ion currents do not decrease too significantly when the HF voltage is turned off, which indicates that the LF controls ion and electron loss from the plasma. Once the LF is turned off, ion and electron currents decrease considerably. There is some lag in the effect of voltage turn-off on electrode currents due to species inertia.

We next focus on the transition when the voltages are turned on. We first consider the situation when the LF voltage is turned on earlier than the HF voltage. In this case, the $\Delta\tau_{on}$ period has a small impact on $n_e$ and $T_e$. As shown in Fig. 12, at the instance when the LF voltage is turned on, $n_e$ and $T_e$ decrease more quickly compared to synchronous pulsing for the range of $\Delta\tau_{on}$ considered in this study. Once the HF voltage is turned on, plasma density quickly builds up and not much effect of negative $\Delta\tau_{on}$ can be observed. The effect of negative $\Delta\tau_{on}$ on the spatiotemporal variation of $n_e$ is shown in Fig. 13. These results show that the sheath expands when the LF voltage is turned on but without any rise in bulk plasma density. This indicates that the LF voltage leads to higher losses than production during this brief period. Electron and ion currents to the electrode are plotted in Fig. 14 for $\Delta\tau_{on} = -5$ µs. These results show that at the instance the LF voltage turns on, the wall current rises sharply indicating high charge losses to the wall. No significant plasma production is apparent from the results in Fig.

13 until the HF voltage is turned on. After the initial spike in currents which is due to the loss of the electrons and ions from the sheath region, the wall current decreases until the HF voltage turns on again.

The last situation we consider is when the LF voltage is turned on after the HF voltage ($\Delta\tau_{on} > 0$). The spatially-averaged $n_e$ and $T_e$ for several positive $\Delta\tau_{on}$ are shown in Fig. 15. Once the HF voltage is turned on, plasma production starts and $n_e$ increases sharply. The electrons also get more energetic. The rate of $n_e$ increase is significantly large until the LF voltage is turned on. The LF voltage increases the loss of electrons and ions from the plasma, thus decreasing the rate at which the plasma density builds up. Positive $\Delta\tau_{on}$ therefore has strong influence on how quickly we can ramp up to the steady-state plasma during the voltage on phase. The spatially-resolved $n_e$ as a function of time is plotted in Fig. 16 for positive $\Delta\tau_{on}$. One can observe here the rapid rise in $n_e$ after the HF voltage is turned on, but also see that the sheaths are thin until the LF voltage is turned on. The electron and ion current to the electrode are shown in Fig. 17 when $\Delta\tau_{on} = 5$ μs. After the HF voltage turns on, there is a brief spike in electron current, which is due to the electrons that leave the region where the sheath forms. The currents remain small afterwards until the LF voltage is turned on.

## 4. CONCLUSIONS

Dual frequency capacitively coupled plasma are widely used for thin film etching and deposition in the semiconductor manufacturing. In these plasmas, the high and low frequency RF sources are meant to provide individual control over plasma specie density and ion energy distribution, respectively. These DF CCPs are often pulsed to enable even finer control over ion / neutral fraction and ion energy. In this work, synchronous rf pulsing and several variations (with phase lag between the 2 RF sources) are studied using a particle-in-cell Monte Carlo plasma model. We found that the small delays when the RF sources are turned on or off have a strong impact on major plasma characteristics such as electron density and temperature. For example, if the low-frequency (LF) source is turned off a few µs before (or after) the high-frequency source, the electron density during the off-state is significantly higher (or lower) due to the frequency coupling effect. Similarly, turning on the LF source a few µs after the HF voltage in a sharp increase in the plasma density when the RF sources are turned on. The voltage ramping rate during source on or off has minimal impact on the plasma properties. One can use these delays to control important plasma properties more precisely in pulsed plasmas. This sensitivity introduces challenges for chamber matching and process repeatability, which need to be considered while designing pulsed plasma processes.

# Figure Captions

1. Geometry and HF and LF pulsed rf voltage source for the plasma simulations

2. Schematic diagram of the pulsed CCP configurations of the HF and LF pulsed rf voltage source

3. Steady-state (a) electron density $n_e$ (b) electron temperature $T_e$, and (c) potential between electrode gap over a pulse period. These results correspond to the base case with synchronous pulsing.

4. Steady-state, spatially averaged (a) electron density $n_e$ (b) electron temperature $T_e$, corresponding to the base case and 'base case with ramp'.

5. Electron and ion wall current to the HF electrode at z = 0 over a pulse period. These results correspond to base case.

6. Steady-state, spatially averaged (a) electron density $n_e$ (b) electron temperature $T_e$, corresponding to base case and staggered pulsing cases St-A-2.5 and St-A-5.0 in Table 1.

7. Steady-state (a) electron density $n_e$ between electrode gap over a pulse period. These results correspond to staggered pulsing cases St-A-2.5 and St-A-5.0 in Table 1.

8. Electron and ion wall current to the HF electrode at z = 0 over a pulse period. These results correspond to the case St-A-5.0 in Table 1.

9. Steady-state, spatially averaged (a) electron density $n_e$ (b) electron temperature $T_e$, corresponding to the base case and pulsing cases St-B-2.5 and St-B-5.0 in Table 1.

10. Steady-state (a) electron density $n_e$ between electrode gap over a pulse period. These results correspond to the pulsing cases St-B-2.5 and St-B-5.0 in Table 1.

11. Electron and ion wall current to the HF electrode at $z = 0$ over a pulse period. These results correspond to the case St-B-5.0 in Table 1.

12. Steady-state, spatially averaged (a) electron density $n_e$ (b) electron temperature $T_e$, corresponding to the base case and the pulsing cases St-C-2.5 and St-C-5.0 in Table 1.

13. Steady-state (a) electron density $n_e$ between electrode gap over a pulse period. These results correspond to the pulsing cases St-C-2.5 and St-C-5.0 in Table 1.

14. Electron and ion wall current the HF electrode over a pulse period. These results correspond to the case St-C-5.0 in Table 1.

15. Steady-state, spatially averaged (a) electron density $n_e$ (b) electron temperature $T_e$, corresponding to the base case and the pulsing cases St-D-2.5 and St-D-5.0 in Table 1.

16. Steady-state (a) electron density $n_e$ between electrode gap over a pulse period. These results correspond to the pulsing cases St-D-2.5 and St-D-5.0 in Table 1.

17. Electron and ion wall current to the HF electrode at $z = 0$ over a pulse period. These results correspond to the case St-D-5.0 in Table 1.

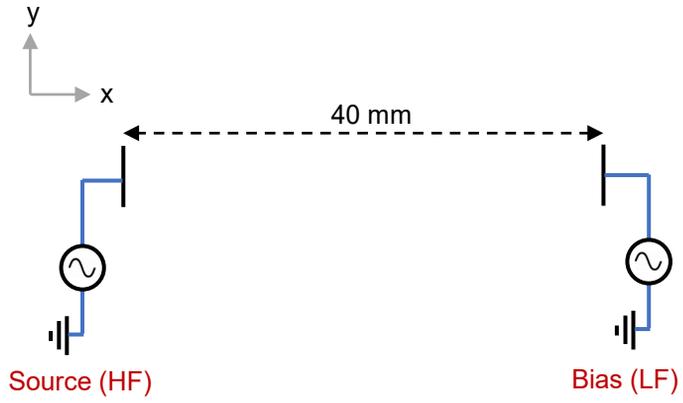

Fig. 1.

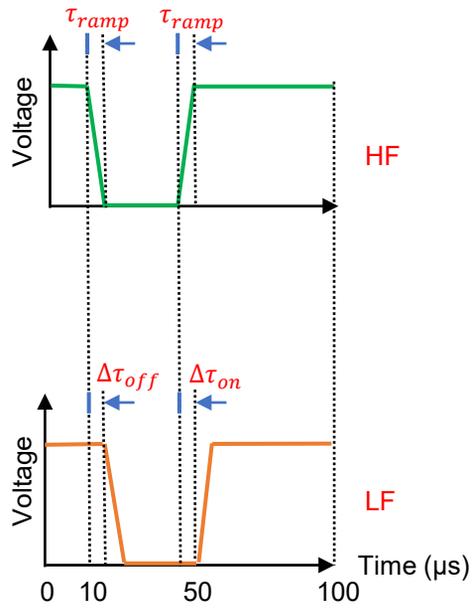

Fig. 2.

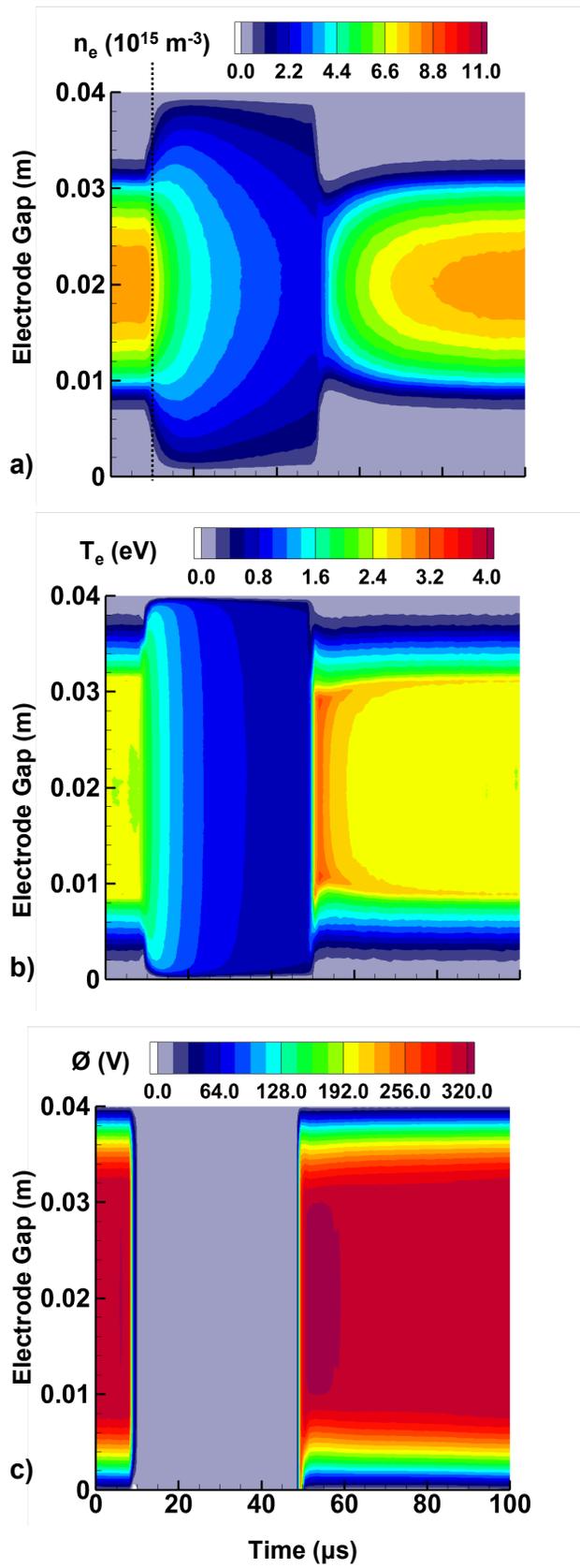

Fig. 3.

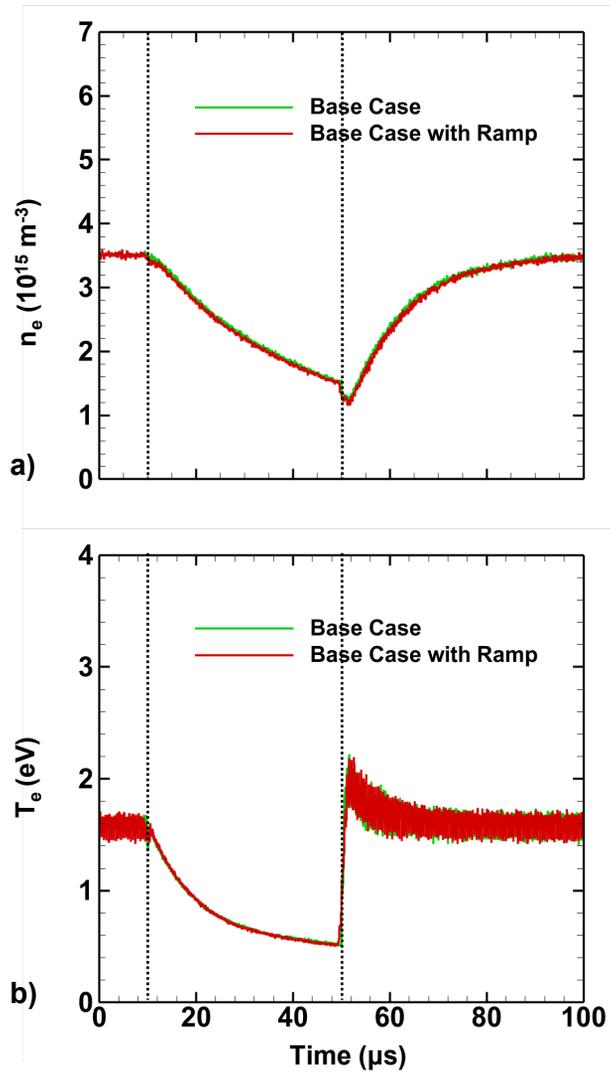

Fig. 4.

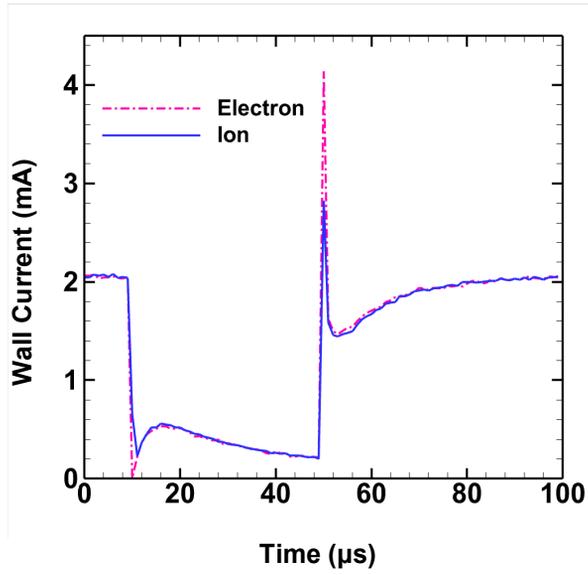

Fig. 5.

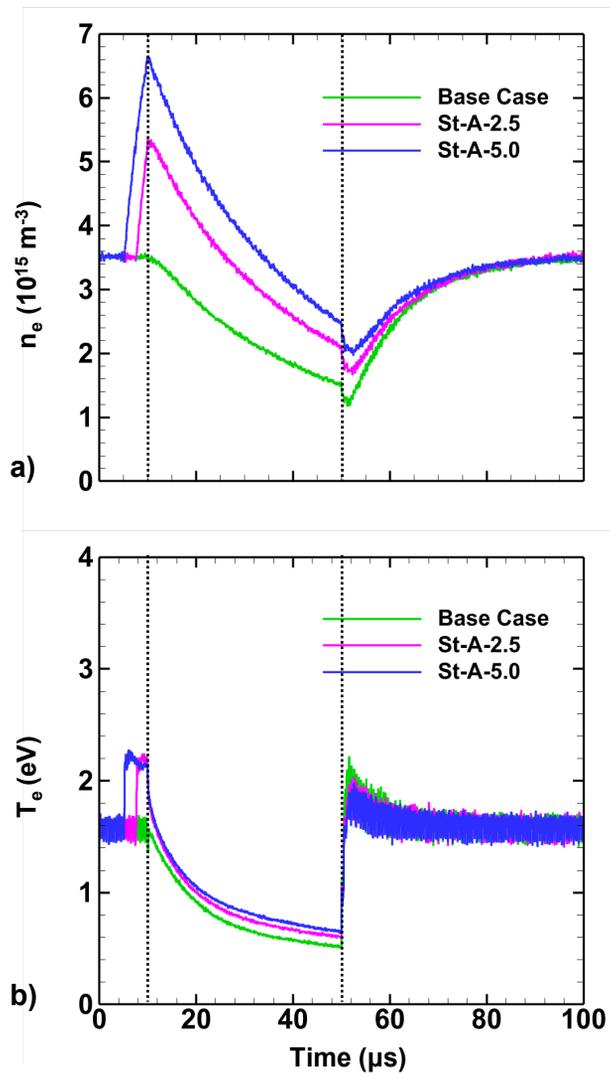

Fig. 6.

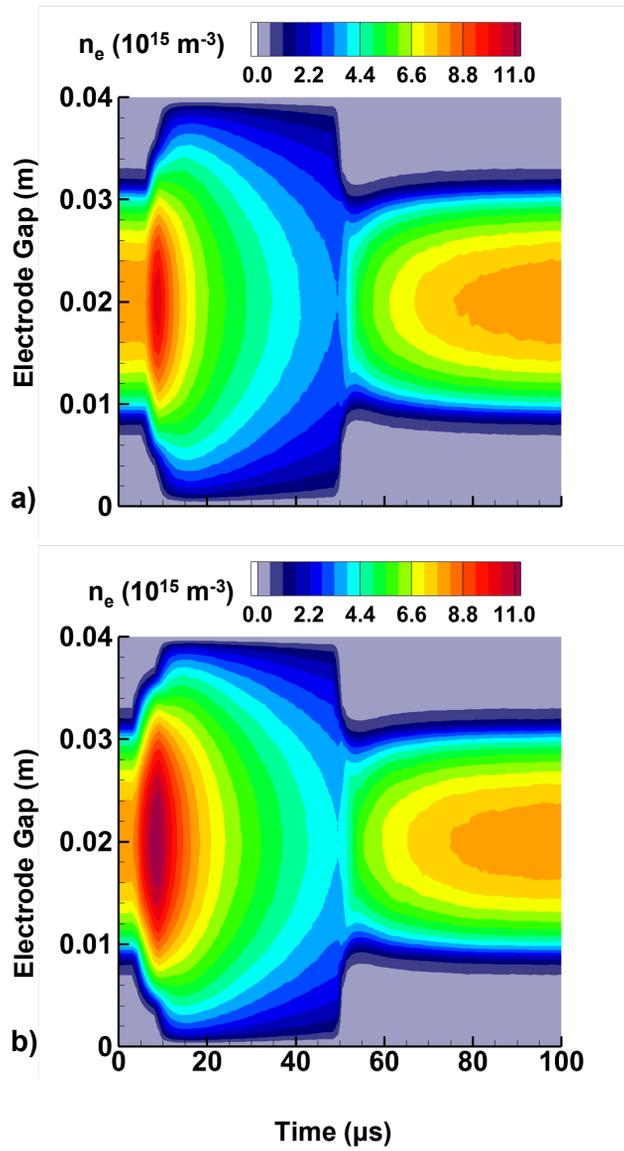

Fig. 7.

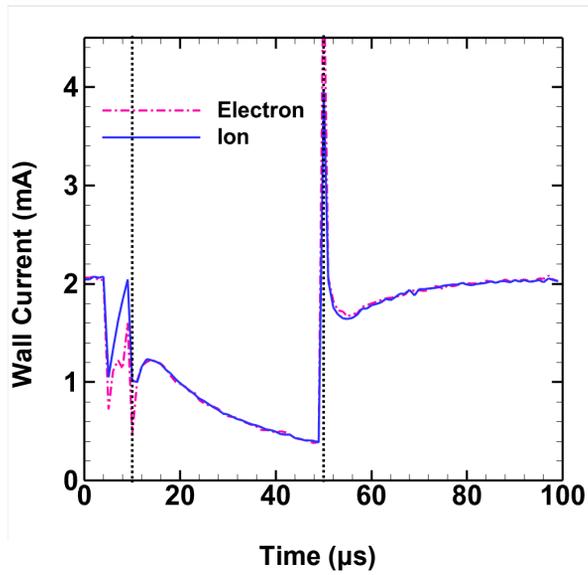

Fig. 8.

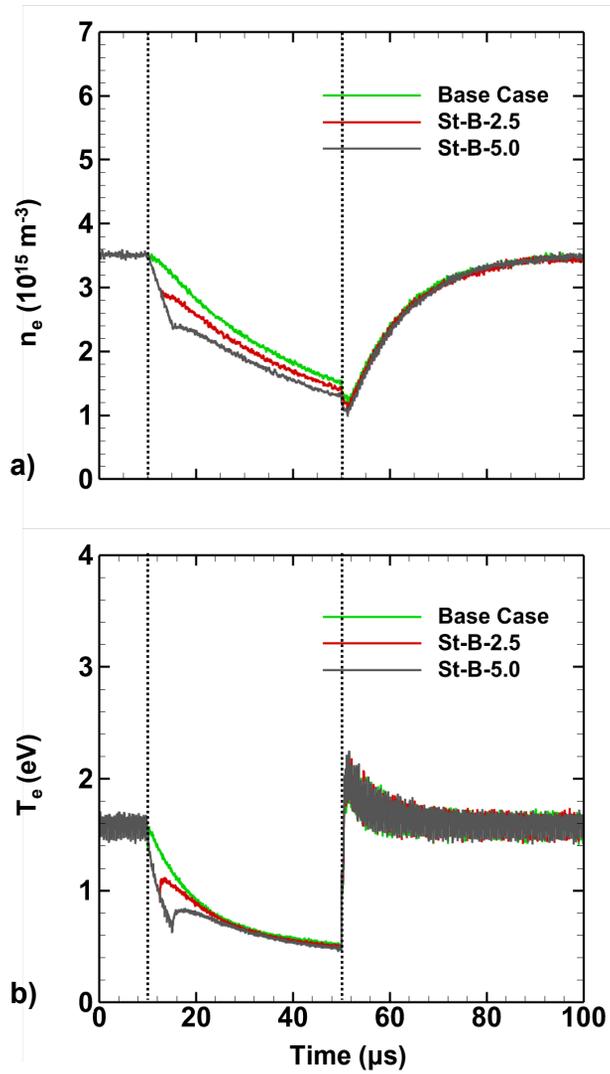

Fig. 9.

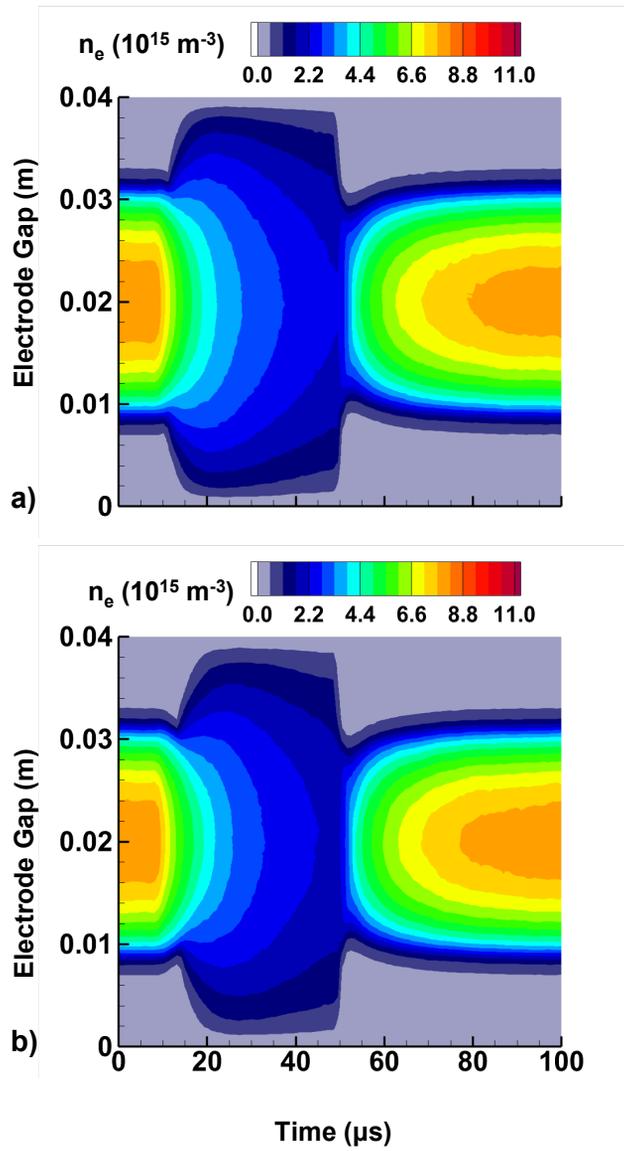

Fig. 10.

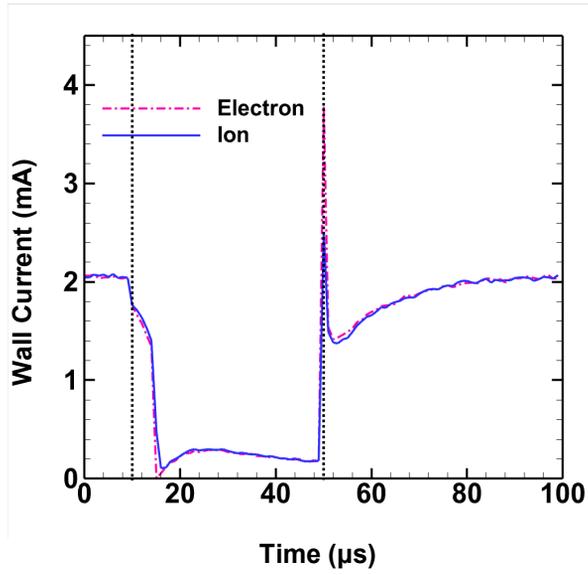

Fig. 11.

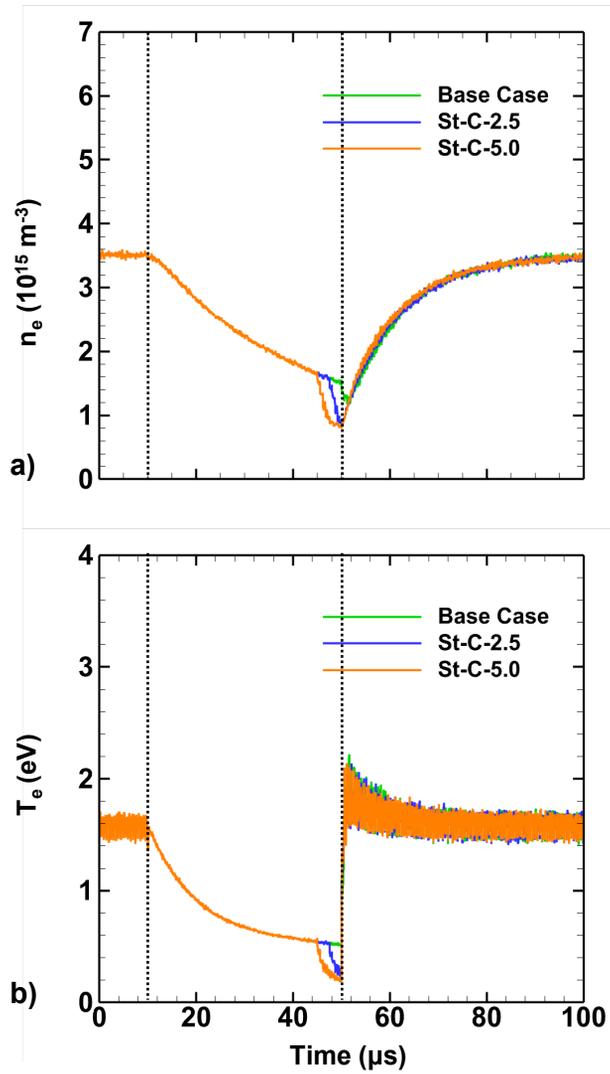

Fig. 12.

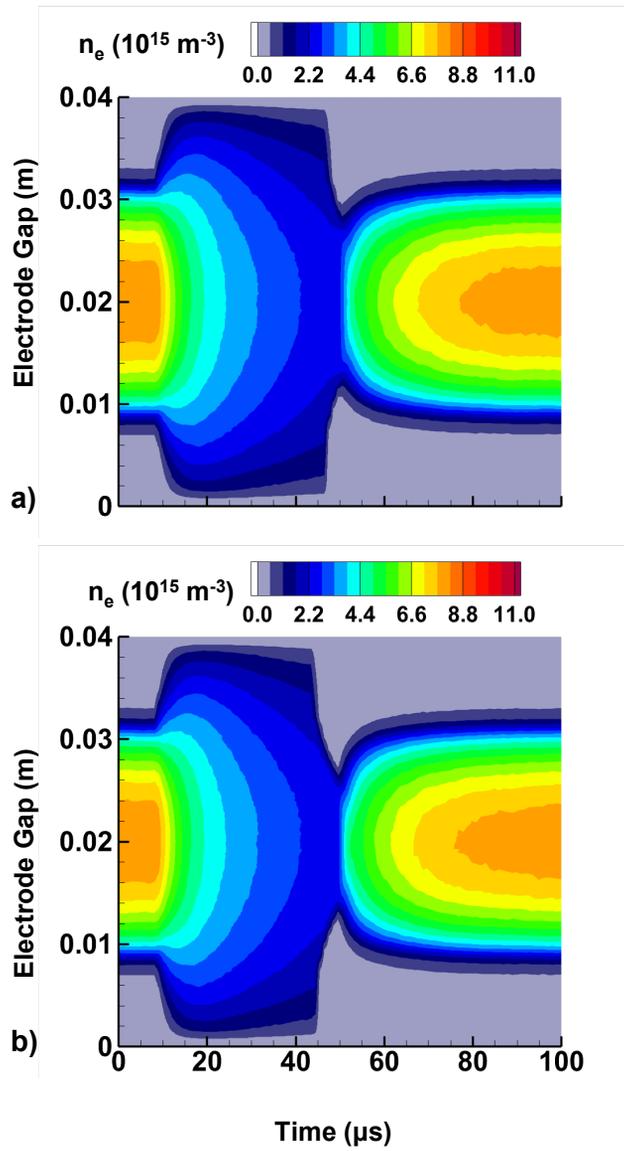

Fig. 13.

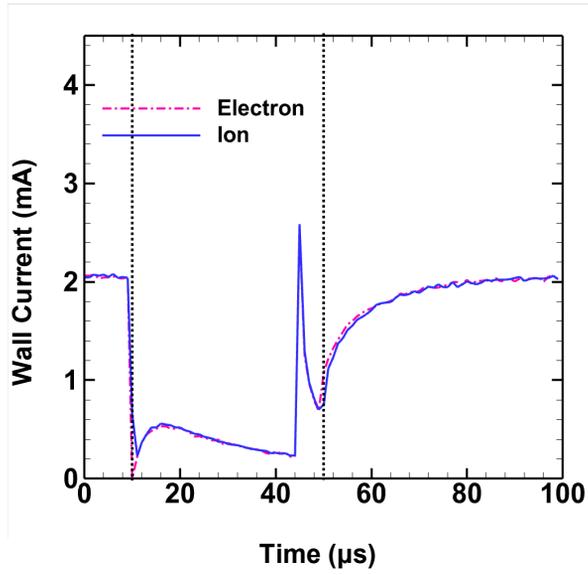

Fig. 14.

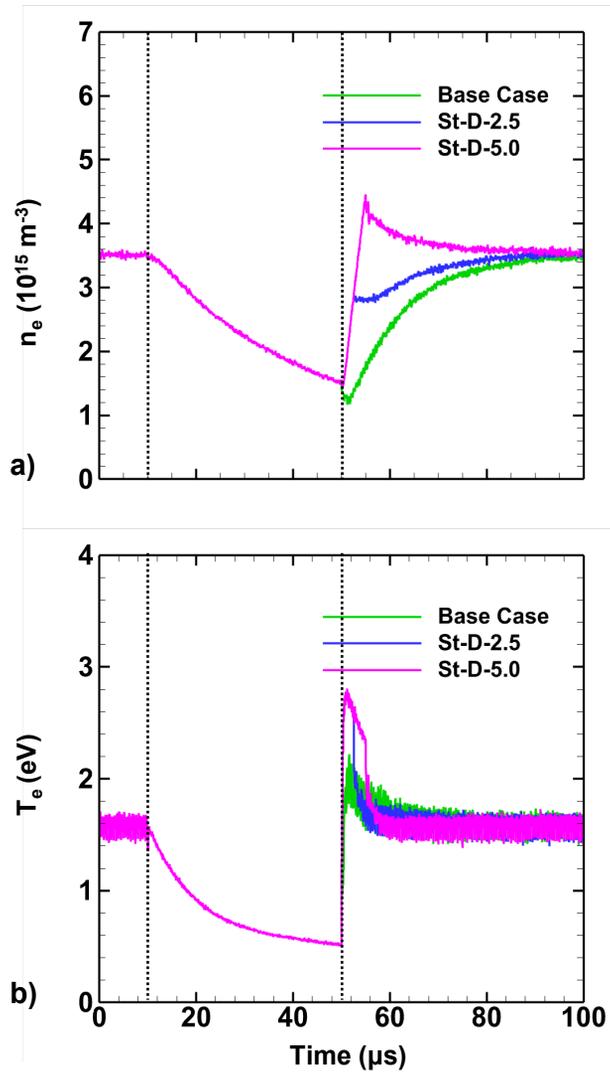

Fig. 15.

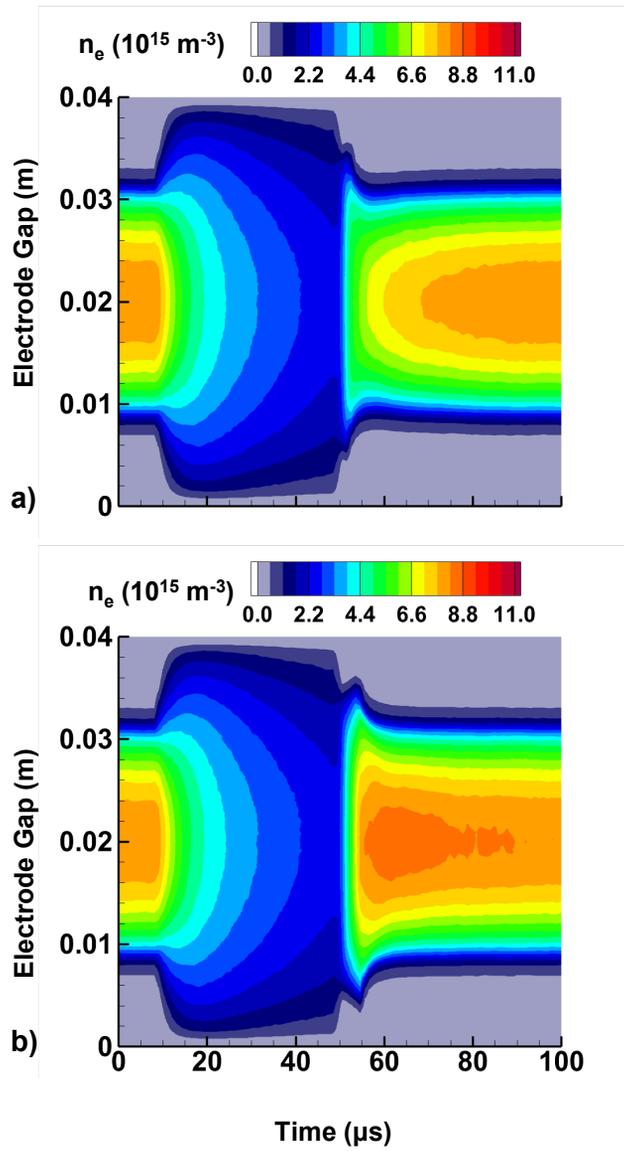

Fig. 16.

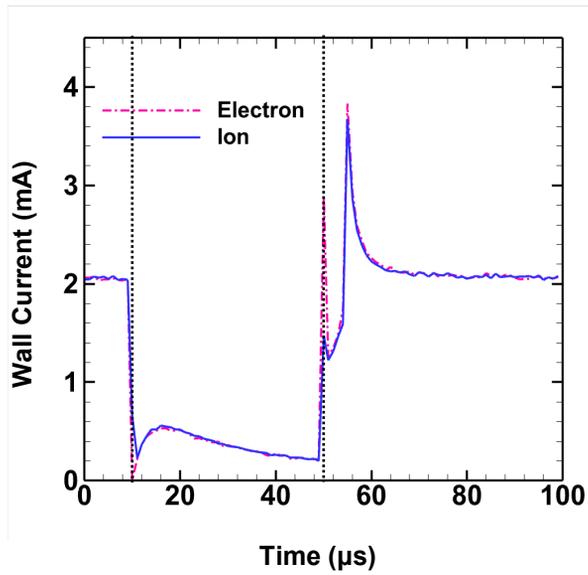

Fig. 17.